\documentclass[prb,aps,twocolumn,superscriptaddress,longbibliography]{revtex4-2}

\usepackage{graphicx,amsfonts,times,bm,amsmath,verbatim,color,array}
\graphicspath{{figs/}}

\usepackage{float}
\usepackage{epsfig}
\usepackage{hyperref}
\hypersetup{
	colorlinks=true,final=true,
	linkcolor=blue,
	citecolor=blue,
	filecolor=blue,
	urlcolor=blue,
}

\begin{document}
	
	\title{Probing mirror anomaly and classes of Dirac semimetals with circular dichroism}
	
	\author{Abhirup Roy Karmakar}
	\affiliation{Department of Physics, Indian Institute of Technology Kharagpur, W.B. 721302, India}
	\author{S. Nandy}
	\affiliation{Department of Physics, University of Virginia, Charlottesville, VA 22904, USA}
	\author{G. P. Das}
	\affiliation{Department of Physics, Indian Institute of Technology Kharagpur, W.B. 721302, India}
	
	\affiliation{Dept. of Metallurgical and Materials Engineering, Indian Institute of Technology Kharagpur, W.B. 721302, India}
	
	\author{Kush Saha}
	\affiliation{School of Physical Sciences, National Institute of Science Education and Research, Jatni, Odisha 752050, India}
	\affiliation{ Homi Bhabha National Institute, Training School Complex, Anushakti Nagar, Mumbai 400094, India}
	
\begin{abstract}
		 We theoretically investigate the optical activity of three dimensional Dirac semimetals (DSMs) using circular dichroism (CD). We show that DSMs in the presence of a magnetic field in any one of the mirror-symmetric planes of the materials exhibit a notable dichroic behavior. In particular, for different orientations of the light field with respect to the mirror-symmetric plane, the CD in type-II DSMs can detect the presence of mirror anomaly by showing sharply distinct patterns at the mirror-symmetric angle. Interestingly, we find that the CD can also distinguish type-II DSMs having only one Dirac point at a time-reversal invariant momentum from type-I DSMs with a pair of Dirac points on the rotation axis of the crystals.
\end{abstract}
	
	\maketitle

\section{Introduction}\label{introduction}
	
Nearing a decade of their theoretical and experimental advances, the three dimensional (3D) Dirac semimetals (DSMs) continue to evolve as promising  materials  for studying several unusual phenomena in condensed matter physics \cite{hasan_colloquium_2010, wan_topological_2011, yang_quantum_2011, xu_chern_2011,neupane_observation_2014, burkov_weyl_2011, xu_discovery_2015, young_dirac_2012, liu_discovery_2014, wang_dirac_2012, wang_three-dimensional_2013} such as large magnetoresistance \cite{xiong_evidence_2015, he_quantum_2014, li_negative_2016, wang_quantum_2017, shekhar_extremely_2015}, giant diamagnetism \cite{wang_dirac_2012}, oscillating quantum spin Hall effect \cite{wang_dirac_2012, burkov_z_2016} \cite{liang_ultrahigh_2015}, etc. Additionally, they can host topological insulators and Weyl metals in the presence and/or absence of certain symmetries such as time-reversal (TR) symmetry, inversion symmetry (IS), and crystalline uniaxial rotational symmetry.  Based on these symmetries, DSMs can further be classified as type-I and type-II \cite{yang_classification_2014,young_dirac_2012}. In type-I DSMs, Dirac points generally appear in pairs on the rotation axis and they are protected by crystalline symmetry. In contrast, type-II DSMs contain a single Dirac point at the time-reversal-invariant momentum (TRIM) on the rotation axis. Na$_{3}$Bi \cite{liu_discovery_2014,Xu294,kushwaha_bulk_2015} and Cd$_3$As$_2$\cite{neupane_observation_2014,he_quantum_2014,liang_ultrahigh_2015,liu_stable_2014,jeon_landau_2014} compounds are found to show type-I behavior, while TlBi(S$_{1-x}$Se$_{x}$)$_2$\cite{sato_unexpected_2011}, (Bi$_{1-x}$In$_x$)$_2$Se$_3$\cite{brahlek_topological-metal_2012}, and ZrTe$_5$\cite{li_chiral_2016,chen_magnetoinfrared_2015} are confirmed to be type-II Dirac semimetals. We note that there is another classification of DSMs based on the tilting of Dirac nodes \cite{sadhukhan_novel_2020, yan_lorentz-violating_2017, burkov_topological_2016}. However, in the present study, we focus on the symmetry based type-I and type-II DSMs without tilting as mentioned above.

Although these two types of DSMs differ in underlying symmetries, apparently there are no as such observable properties that are significantly different in these two class of DSMs. Since both of them possess chiral anomaly, it is expected to observe negative longitudinal magnetoresistance or planar Hall effect in both of these DSMs in the presence of electric and magnetic fields. In addition to the chiral anomaly, DSMs have been shown to possess mirror anomaly which is manifested in a step-like anomalous Hall conductivity (AHC) as a function of polar angle of the applied magnetic field perpendicular to the mirror-symmetric plane, resembling the AHC due to the parity anomaly in 2D systems with massive Dirac fermions \cite{burkov_mirror_2018}. However, the step-like behavior is found to be broadened in the presence of an additional perturbation that breaks mirror symmetry. Accordingly, the Hall measurement can be thought of as a key observable to infer the presence or absence of mirror anomaly in type-II DSMs. However, such measurements are always limited by impurities or it may have contribution from orbital effect, which in turn may lead to the deviation from the step-like behavior. Thus the anomalous Hall measurement may not be an ideal probe to discriminate between type-II DSMs with or without mirror symmetry. In addition, very recently, it has been shown that this step-like feature is not generic to all DSMs. In type-I Dirac semimetals, the step-like AHC is smoothen out with the polar angle across the mirror-symmetric plane \cite{nandy_mirror_2019} even in the presence of mirror symmetry. Therefore the types of DSMs discussed herein may also not be distinguished experimentally with the help of Hall measurement.     
Hence, it is desirable to find easily accessible diagnostic tools to discriminate  DSMs with or without mirror symmetry as well as their types. This may eventually help us understanding the optical responses of DSMs with different underlying symmetries and their potential applications in opto-electronic devices. 
    
In view of that, we theoretically study the optical activity of these two types of DSMs via circular dichorism (CD) which deals with differential absorption of left- and right-circularly polarized light of materials under consideration. It has already been shown that the chiral anomaly of Weyl semimetals with pairs of Weyl nodes with opposite chiralities can be probed by the CD\cite{hosur_2014}. In contrast, the CD provides vanishing result for pristine DSMs since the Weyl nodes with opposite chiralities coincide in energy and momentum space. The reason for the null results can further be understood from the connection between the CD and Berry curvature. The momentum resolved circular dichorism is found to be proportional to the component of the Berry curvature along the direction of incident light\cite{liu_circular_2018}. Thus for gapless DSMs, the Berry curvature is zero everywhere in momentum space except at the gapless point. This leads to the null CD. However, DSMs under symmetry breaking may give rise to finite Berry curvature, hence nonzero CD. To verify this, we apply magnetic field along any one of the mirror-symmetric planes of the DSMs. With this, we find that the optical responses of both types of DSMs depend on the orientation of the light field. In particular, the CD due to the light field along $x$ direction differs from the CD with the light field along $y$ direction for both types of DSMs. Moreover, we find that the presence or absence of mirror symmetry in type-II DSMs is well manifested in the CD, specially at the mirror-symmetric angle. We furthermore show that there is a sharp contrast between the CD in type-I and type-II DSMs. 
When the light field is incident along the $x$ direction, at the mirror-symmetric angle, the CD diverges at the zone center in type-II DSMs, whereas it is found to be zero in type-I DSMs. 
Additionally, the CD is obtained to be an odd function of each component of momentum for type-II, but it differs for type-I. Indeed, these signatures are key to identify different types of DMSs with or without mirror symmetry, which in turn may give us indirect hint on if a DSM can possess mirror anomaly.  
    
The rest of the paper is organized as follows. In Sec.~\ref{general_theory} we provide a general framework for the CD, relating it to the Berry curvature. In Sec.~\ref{type-ii_dirac_semimetal}, we discuss the Berry curvature and corresponding CD in type-II DSMs for different orientations of the light field. We compare results between the cases with and without mirror symmetry. This is followed by Sec.~\ref{type-ii_dirac_semimetal} where we only focus on the type-I DSMs with mirror symmetry and compare the results for two different directions of the applied light field. We also compare the results with the type-II DSMs. We conclude in Sec.~\ref{conclusion} with a discussion on possible applications.

\section{General theory: Optical transition}\label{general_theory}

We begin by reviewing light-induced optical activity in a two-band model, particularly focusing on the Berry-curvature dependent interband transition probabilites. The light field couples to the electron of a material in two ways: a) orbital coupling where momentum $\bm k$ is replaced by $\bm k-e \bm A$ due to minimal coupling and b) Zeeman coupling via $\bm \nabla\times \bm A$, where $\bm A$ is the vector potential due to the light field. Thus the total Hamiltonian of light-electron interaction reads off 
\begin{align}
\mathcal H_{le}=-e\, \bm v^{k}.\,\bm A\,+\,g_s\mu_B (\bm \nabla\,\times\, \bm A)\cdot\bm \sigma,
\end{align}	
where  $e$ is the charge of electron, $m_e$ is the mass of the electron, $g_s$ is the Lande-$g$ factor, $\mu_B$ is the Bohr magneton and  $\bm v^{k}=\frac{1}{\hbar}\partial \mathcal{H}/\partial {\bm k}$ with $\mathcal H$ being the noninteracting Hamiltonian of the system in consideration. The absorption and emission of the photon lead to particle-hole excitations which in turn lead to optical transition between energy bands. Since the term associated with the orbital coupling gives rise the Berry-curvature dependent interband transition, one can safely neglect \textit {Zeeman} coupling with the light field. In addition, for a typical Dirac material with Fermi velocity $v_F=10^5\,m/s$ and photon energy $0.1$ eV, the ratio between orbital coupling and \textit{Zeeman} coupling turns out to be $~10^{5}$.
Thus we mainly focus on the orbital part of $\mathcal H_{le}$. 

We assume light propagating along $z$ direction with a vector potential $\bm A=A_0(\bm{\hat{e}_x}\pm i\,\bm{\hat{e}_y})$, where $\bm{\hat{e}_x}, \bm{\hat{e}_y}$ are unit vectors along $x$ and $y$ directions, respectively; $+$ denotes right circularly polarized light and $-$ denotes left circularly polarized light. With this, the orbital part of $\mathcal H_{le}^{\rm}$ can be written as
\begin{align}
\mathcal H_{le}^{\rm orb\pm}=-e\,A_0\, (v^k_x\pm i\, v^k_y).
\end{align}
Accordingly, for a two-band model, the optical transition matrix is defined as 

\begin{equation}\label{trans_mat}
\mathcal P_{\pm}(\bm k) = \mathcal P_{x} \pm i\, \mathcal P_{y} = \langle c|H_{le}^{\rm orb\pm} |v \rangle,
\end{equation}
where $|c \rangle$ and $|v \rangle$ are conduction and valence bands, respectively. 
Then the $k-$resolved circular dichroism is given by \cite{yao_valley-dependent_2008}
\begin{equation}\label{cd-tm}
\eta(\bm k) \equiv \frac{|\mathcal{P}_+(\bm k)|^2 - |\mathcal{P}_-(\bm k)|^2}{|\mathcal{P}_+(\bm k)|^2 + |\mathcal{P}_-(\bm k)|^2}.
\end{equation}
The numerator of $\eta$ can further be expressed in terms of the valence band Berry curvature (${\bf \Omega}$) of a two-band model as follows: 
\begin{align*}
|\mathcal{P}_+(\bm k)|^2 &- |\mathcal{P}_-(\bm k)|^2 = 2i(\mathcal{P}_{x}^*\mathcal{P}_{y}-\mathcal{P}_{y}^*\mathcal{P}_{x}) \\
&=2i \frac{e^2\,A_0^2}{\hbar^2} (\left< v|\partial\mathcal H/\partial k_{x}| c \right>\left< c|\partial\mathcal H/\partial k_{y}| v \right>\\
& \qquad\quad - \left< v|\partial\mathcal H/\partial k_{y}| c \right>\left< c|\partial\mathcal H/\partial k_{x}| v \right>)\\
&= \frac{2\,e^2\,A_0^2 (\epsilon_{c}-\epsilon_{v})^2}{\hbar^2} \bm{\Omega}\cdot \bm{\hat{z}}
\end{align*}

For an arbitrary direction ($ \bm{\hat{n}} $) of the incident light, we can generalize $\eta({\bf k})$ by rotating the coordinate frame accordingly to obtain 
\begin{align}\label{eta}
\eta(\bm k) &= \frac{e^2A_0^2\, (\epsilon_{c}-\epsilon_{v})^2}{\hbar^2} \frac{\bm{\Omega}\cdot \bm{\hat{n}}}{\mathcal{P}_t}, 
\end{align}
where $\mathcal{P}_t=|\mathcal{P}_+(\bm{k^0})|^2 + |\mathcal{P}_-(\bm{k^0})|^2$ and $\bm{k^0}=R\,{\bm k} $ with $R$ being the usual rotation matrix. 
	
We next aim to investigate $\eta$ in type-I and type-II Dirac semimetals in the presence of symmetry breaking, considering lowest conduction and highest valence bands. The simple form of the low-energy Hamiltonian of type-II DSMs allows us to find analytical expressions for the CD for some specific directions, while type-I is needed to be solved numerically due to complexities in the structure of the Hamiltonian. We therefore first discuss the CD for type-II followed by the same for type-I.

\section{Circular Dichroism in Type-II Dirac Semimetal}\label{type-ii_dirac_semimetal}
	
The linearized low energy model Hamiltonian for a type-II DSM is given by \cite{burkov_mirror_2018},
\begin{equation}\label{h-ii}
\mathcal{H}^{\textrm{II}}(\bm k) = v\,(-k_x \tau_z \sigma_y + k_y \tau_z \sigma_x + k_z \tau_y ),
\end{equation}
where $ v$ is the quasiparticle velocity, $\sigma$'s and $\tau$'s are the Pauli matrices for spin and orbital, respectively. Note that, we work here with $\hbar=1$. This Hamiltonian in Eq.~(\ref{h-ii}) has only one Dirac node at the time-reversal invariant momentum (TRIM) $\Gamma$. Eq.~(\ref{h-ii}) commutes with the Mirror symmetry operator $M=i\sigma_{x}$ at $ k_x=0 $ plane:
\begin{equation}\label{h_comm}
[\mathcal{H}^{\textrm{II}}(k_x=0) , M]=0.
\end{equation}
Thus the Hamiltonian possesses mirror symmetry in $yz$ plane. The same is true for $xz$ plane as well. In addition, the Hamiltonian in Eq.~(\ref{h-ii}) is invariant under time reversal ($T=i\sigma_y \mathcal {K}$, where $\mathcal{K}$ is complex conjugation operator) and inversion ($I=i\sigma_y$) symmetry. Consequently, we obtain vanishing Berry curvature which in turn may give rise to null CD (cf. Eq.~\ref{eta}). 

To have non-zero Berry curvature, we consider an {\it external} magnetic field in one of the mirror-symmetric planes (say, $xz$). This gives \textit{Zeeman} term as

\begin{equation}\label{h-ii_zmn}
	\mathcal{H}^{\textrm{II}}_{zmn}(\theta) = b\cos(\theta)\sigma_z + b\sin(\theta)\sigma_x,
\end{equation}
where $ b $ is the strength of the field and $\theta$ is the angle between $z$-axis and the direction of the magnetic field. Eq.~(\ref{h-ii_zmn}) preserves mirror symmetry but breaks time reversal symmetry which splits the single Dirac node into two Weyl nodes with opposite chiralities on the $z$ axis at $k_z = \pm b/v$ for all values of $\theta$ except at $\theta=\pi/2$. At this value, the valence and conduction bands touch each other along a nodal line in the $yz$ plane, satisfying $ k_y^2 + k_z^2 = b^2/v^2$. This nodal line is found to be protected by the mirror symmetry with respect to the reflections along the $yz$ plane. Note that $\theta=\frac{\pi}{2}$ is a critical angle, namely mirror-symmetric angle where the chiralities of the Weyl nodes on the $z$  axis interchange sign. It turns out that the emergent mirror symmetry at this critical angle is manifested in the CD as will be discussed shortly.

\begin{figure*}
\includegraphics*[width=1.0\linewidth]{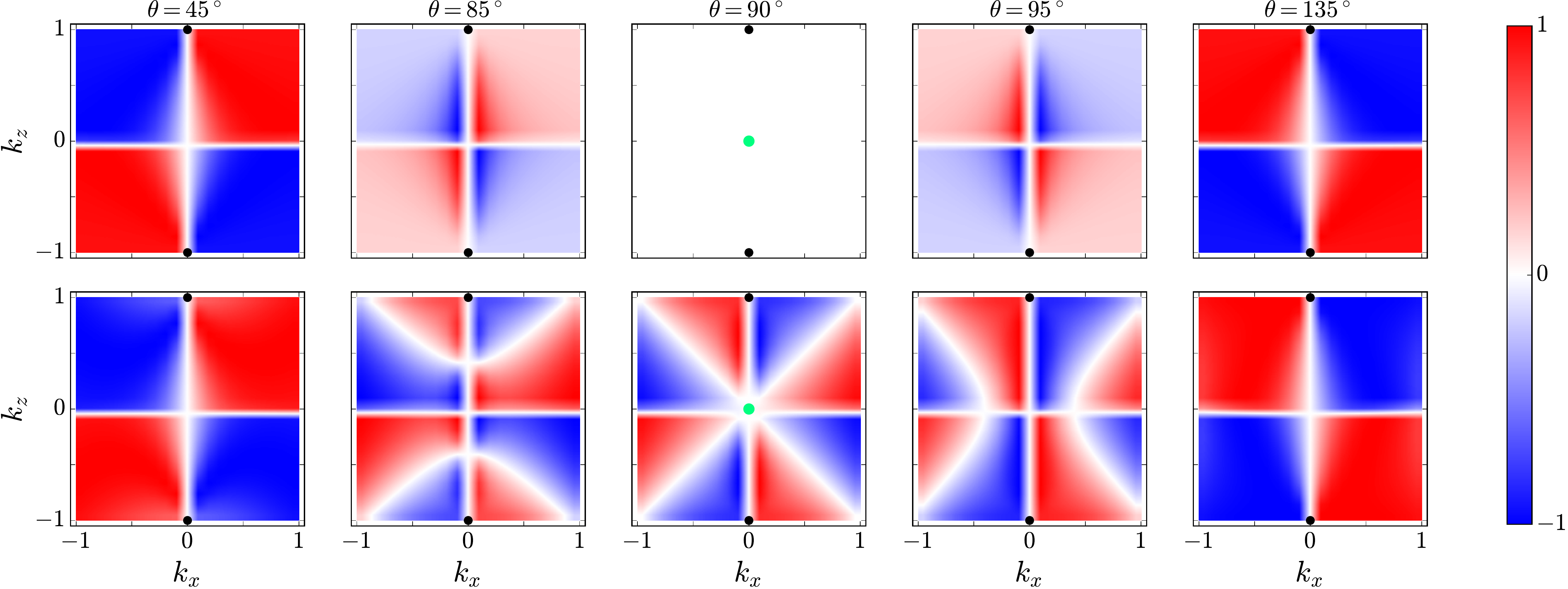}
\caption{Contour plot of circular dichorism, $\eta({\bf k})$ for type-II DSMs with (upper row, $\lambda=0$) and without (lower row, $\lambda=1$) mirror symmetry for different angles of the applied magnetic field in the $xz$ plane. The incident light is taken along the $x$-direction. The \textit{black dots} denote the location of Weyl nodes in momentum space. Evidently, there is a clear distinction in the CD pattern between upper row and lower row at the mirror-symmetric angle, i. e, at $\theta=\pi/2$ for finite momentum. However, $\eta$ diverges for the both the cases at $\bf k=0$ as represented by green dots. Here, $ v/b=1$ in units of $\AA$.}
\label{fig:eta-ii-x}
\end{figure*}
	
Since the momentum resolved CD is proportional to the Berry curvature, it is instructive to find an analytical expression of it to have an intuitive understanding on the CD. A block diagonal form of the the mirror-symmetric Hamiltonian $\mathcal{H}^{\textrm{II}} + \mathcal{H}^{\textrm{II}}_{zmn}$ may help finding such an expression of the Berry curvature.  It turns out that Eq.~(\ref{h-ii}) can be block-diagonalized by rotating the $z$-axis along $B$ followed by   similarity transformations $\sigma_{x,y} \rightarrow \tau_z \sigma_{x,y}$ and $\tau_{x,y} \rightarrow \tau_{x,y} \sigma_z$ \cite{burkov_mirror_2018}. With this, we obtain 
\begin{equation}\label{h-ii_2x2_mirr}
	\mathcal{H}_{2\times 2} = v\, k_y \cos\theta \sigma_x-v\, k_x \sigma_y + p_\pm \sigma_z,
\end{equation}
where, $ p_\pm = b\pm v \sqrt{k_y^2 \sin^2\theta+k_z^2} $.
The $ + $ and $ - $ signs correspond to two blocks of the block-diagonalized Hamiltonian. We focus on the block, associated with the $ - $ sign, that contains the lowest conduction and highest valence bands with energies
\begin{equation}\label{en-ii_mirr}
	\epsilon_{\pm,\bm k} = \pm \epsilon_{\bm k} = \pm \sqrt{v^2k_x^2+v^2k_y^2\cos^2\theta+p_-^2}.
\end{equation}
The corresponding valence band Berry curvature is found to be

\begin{widetext}
\begin{equation}\label{berry_II_mirror}
\bm{\Omega}(\bm k)=
\left\{-\frac{k_x k_z v^3 \cos (\theta )}{2 \epsilon_{\bm k}^3 \sqrt{k_y^2 \sin ^2(\theta )+k_z^2}},-\frac{k_y k_z v^3 \cos (\theta )}{2 \epsilon_{\bm k}^3 \sqrt{k_y^2 \sin ^2(\theta )+k_z^2}}, \frac{v^2 \cos (\theta )}{2 \epsilon_{\bm k}^3} \left(b-\frac{k_z^2 v}{\sqrt{k_y^2 \sin ^2(\theta )+k_z^2}}\right)\right\}.
\end{equation}
\end{widetext}
Note that all three components of the Berry curvature vanish at the mirror-symmetric angle $\theta=\pi/2$. This fact is expected to be reflected in the CD.	

To compare the CD with the case without mirror symmetry, we add a cubic term $ \mathcal{H'}(\lambda) = \frac{\lambda}{2} (k_z^2-k_x^2)k_y \tau_z\sigma_z$ \cite{burkov_mirror_2018} to $ \mathcal{H}^{\textrm{II}} + \mathcal{H}^{\textrm{II}}_{zmn} $. Although the positions of the Weyl points along $k_z$ remain unchanged, instead of $\theta=\pi/2$, they now interchange their chiralities at a critical angle $\theta_c=\cot^{-1}(\delta)$, where $\delta = \lambda b^2/2v^3$.
In addition to these two Weyl points, four additional Weyl points at $(k_y, k_z)\equiv (\pm \frac{b\sin\theta}{v}\sqrt{1-\frac{\cot\theta}{\delta}}, \pm \frac{b}{v} \sqrt{\frac{\cot\theta}{\delta}})$ emerge\cite{burkov_mirror_2018} in the $yz$-plane for $\theta>\theta_c$. These extra Weyl points move slowly towards the \textit{y-axis} and they annihilate with each other at $\theta=\pi/2$. The original two Weyl points retain with their interchanged chiralities.

The Hamiltonian $ \mathcal{H}^{\textrm{II}} + \mathcal{H}^{\textrm{II}}_{zmn} + \mathcal H'(\lambda)$ can be written in a block-diagonalized form using similarity transformations mentioned earlier: 
\begin{equation}\label{h-ii_2x2_no-mirr}
	\begin{split}
	\mathcal H^\lambda_{2\times 2} = [v k_y \cos\theta &- k_y (k_z^2-k_x^2)\lambda/2] \sigma_x \\
	& - v k_x \sigma_y + q_\pm \sigma_z,
	\end{split}
\end{equation}
where
$ q_\pm = b \pm \sqrt{v^2k_z^2 + [v\sin\theta + \cos\theta(k_z^2-k_x^2)\lambda/2]^2k_y^2} $. Diagonalizing Eq.~(\ref{h-ii_2x2_no-mirr}), we obtain the energies for lowest conduction and highest valence bands as
\begin{equation*}\label{en-ii_no-mirr}
	\epsilon_{\pm,\bm k,\lambda} = \pm \sqrt{v^2k_x^2+[v\cos\theta - \sin\theta(k_z^2-k_x^2)\lambda/2]^2k_y^2+q_-^2}.
\end{equation*}
Subsequently, one can easily compute all the three components of the Berry curvature which turns out to be very lengthy to present herein. However, for the sake of further discussions on the CD, we only show two components for two specific limits:
\begin{equation}\label{womirror_Berry}
\Omega_x(k_y=0) = - \frac{k_x k_z v^2 [v \cos\theta + \frac{\lambda}{2}(k_x^2 - k_z^2) \sin\theta]}{2 |k_z| \epsilon_{+,\bm k,\lambda}^3}
\end{equation}
\begin{align}
\Omega_y(k_x=0) = -\frac{k_y\, k_z\, v}{4 \epsilon_{+,\bm k,\lambda}^3(b-q_-)}\left[v\,(2 v^2 + k_y^2 k_z^2\lambda^2)\cos\theta \right.\nonumber\\\left.+(2\,b\,q_- -2\,b^2+(2 k_y^2 +k_z^2) v^2) \lambda \sin\theta\right]
\end{align}

Note that both of these cases exhibit finite Berry curvature at $\theta=\pi/2$ in contrary to the mirror-symmetric case. Hence, these feature is expected to be reflected in the CD. 

We next turn to analyze \textit{circular dichroism} for two distinct cases: (\textit{i})  incoming light in the plane of the applied magnetic field (along $ x $-axis) and (\textit{ii}) incoming light perpendicular to the plane of the applied magnetic field (along $ y $-axis).

\subsection{Light parallel to the plane of magnetic field}
	
Having discussed the Berry curvature for type-II DSMs with and without mirror symmetry, we now investigate $\eta({\bf k})$. For light field along $x$-direction, only $x$-component of $\Omega({\bf k})$  contributes to $\eta({\bf k})$ as evident from Eq.~(\ref{eta}). Since $\Omega_x({\bf k})$ is trivially zero for both the $k_x=0 $ and $k_z=0 $ planes, we concentrate on the $k_y=0$ plane. Together with Eq.~(\ref{eta}) and Eq.~(\ref{berry_II_mirror}), we obtain
\begin{equation}\label{eta-x_ky0}
\eta^{x}(k_x,k_z) = \frac{2\, \epsilon_{\bm k} k_x k_z v \cos \theta }{|k_z| \left(\epsilon_{\bm k}^2 (1+\cos ^2\theta)-p_-^2\right)}.
\end{equation}
The upper row of the Fig.~(\ref{fig:eta-ii-x}) shows contour plot of $\eta^x$ in the $xz$ plane for various $\theta$ in the case of a mirror-symmetric type-II DSM. For $\theta\ll \pi/2$, $\eta^x\rightarrow \pm 1$ except at or in the vicinity of $(k_x,k_z)\equiv 0$. This can be easily understood from Eq.~(\ref{eta-x_ky0}) as follows. We approximate $\cos(\theta)\sim 1$ for $\theta\ll \pi/2$, which in turn simplifies the denominator as $\epsilon_k^2+v^2\, k_x^2$. This further can be approximated as $2\epsilon_k v\,k_x$ in the limit of  $(k_x,k_z)\rightarrow b/v$. We note that $\epsilon_{\bf k}$ is independent of the angle $\theta$ for $k_y=0$ (cf. Eq.~(\ref{en-ii_mirr})). $\eta^x$ starts to deviate as $\theta$ approaches $\pi/2$. At $\theta=\pi/2$, $\eta^x$ diverges at $(k_x,k_z)\equiv (0,0)$ although Berry curvature itself vanishes everywhere in the ($k_x$-$k_z$) plane except at the two Weyl points in this plane. This is due to the fact that the denominator in Eq.~(\ref{eta-x_ky0}) goes as $k_x^2$ at $\theta=\pi/2$, hence $\eta^x\sim 1/k_x.$ 
Notice that $\eta^x$ reverses its pattern as $\theta$ crosses $\pi/2$, corroborating the switching of topological charge of Weyl points across the mirror-symmetric angle. 

\begin{figure}[h]
	\includegraphics[width=0.93\linewidth]{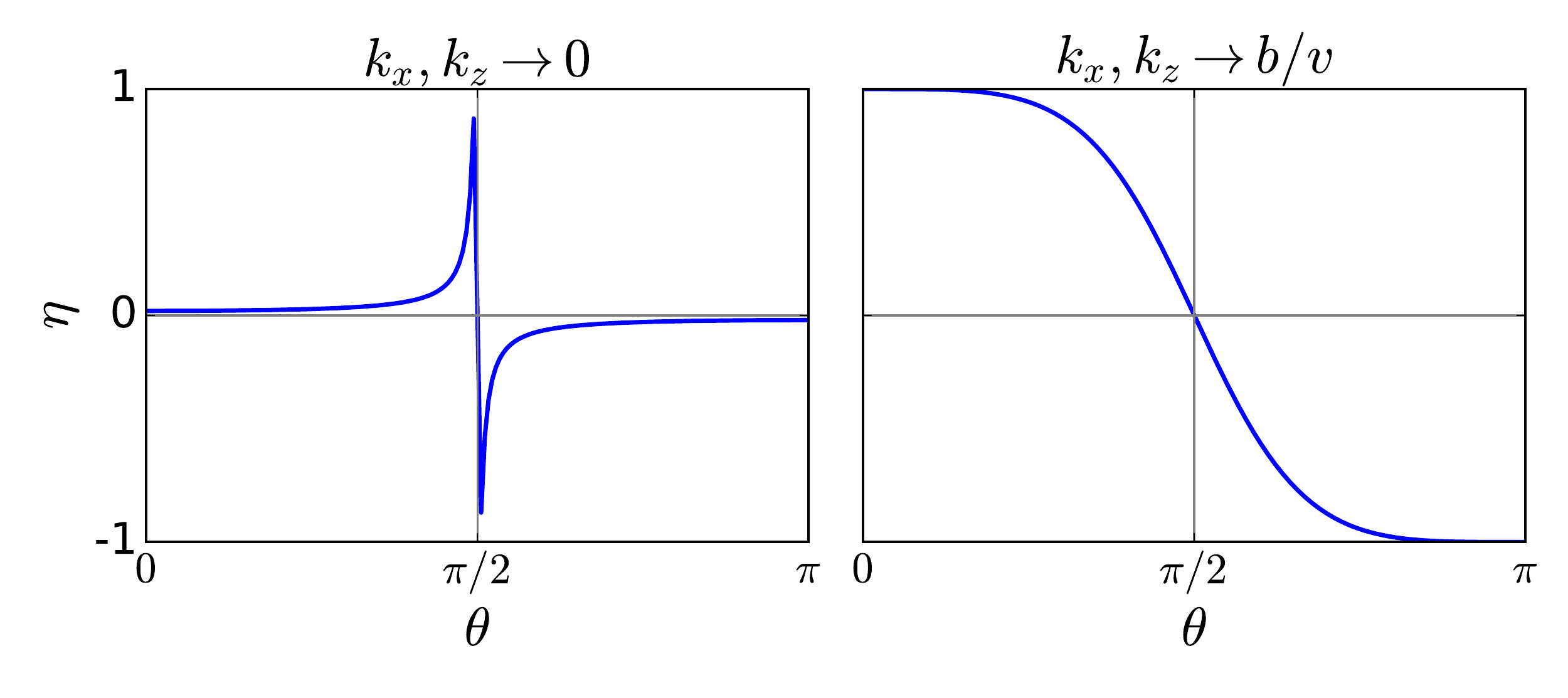}
	\caption{Circular dichroism ($\eta$) as a function of the angle ($\theta$) of the external magnetic field for two extreme limits of $\mathbf{k}$ in a mirror-symmetric type-II DSM. Here, the light is incident along $x$-direction. The parameters used here are the same as mentioned in the caption of Fig.~1.}
	\label{fig:dsm_ii_xz_x_theta}
\end{figure}

To substantiate the behavior of $\eta^x$ in Fig.~(\ref{fig:eta-ii-x}), we show the variation of $ \eta^{x} $ with $\theta$ in Fig~(\ref{fig:dsm_ii_xz_x_theta}) at two different limits  $(k_x,k_z)\rightarrow 0$ and $(k_x,k_z)\rightarrow b/v$. Clearly, in the vicinity of $(k_x,k_z)\rightarrow 0$, $\eta^x$ is found to be negligibly small for all $\theta$ except at $\theta=\pi/2$ where it diverges. On the other hand, $\eta^x\sim 1$ for large $(k_x,k_z)$ for almost all $\theta$ except at $\theta=\pi/2$ where it vanishes. The reason for this behavior can be easily traced back to the $\theta$ dependent numerator i. e., Berry curvature and the denominator $(\mathcal{P}_t)$ of $\eta^x$ (cf. Eq.~\ref{eta}). For  $(k_x, k_z) \rightarrow 0$, $\Omega$ goes as $\cos(\theta)$, whereas $P_t$ goes as $\cos^2\theta$, hence $\eta^{(x)}\sim \frac{1}{\cos\theta}$. In the limit of $(k_x, k_z) \rightarrow b/v$, $\Omega$ goes as $\cos(\theta)$ as before but $P_t$ goes as $1+\cos^2\theta$, which in turn leads to $\eta\sim \frac{\cos(\theta)}{1+\cos^2(\theta)}$.

\begin{figure*}
	\includegraphics[width=1.0\textwidth]{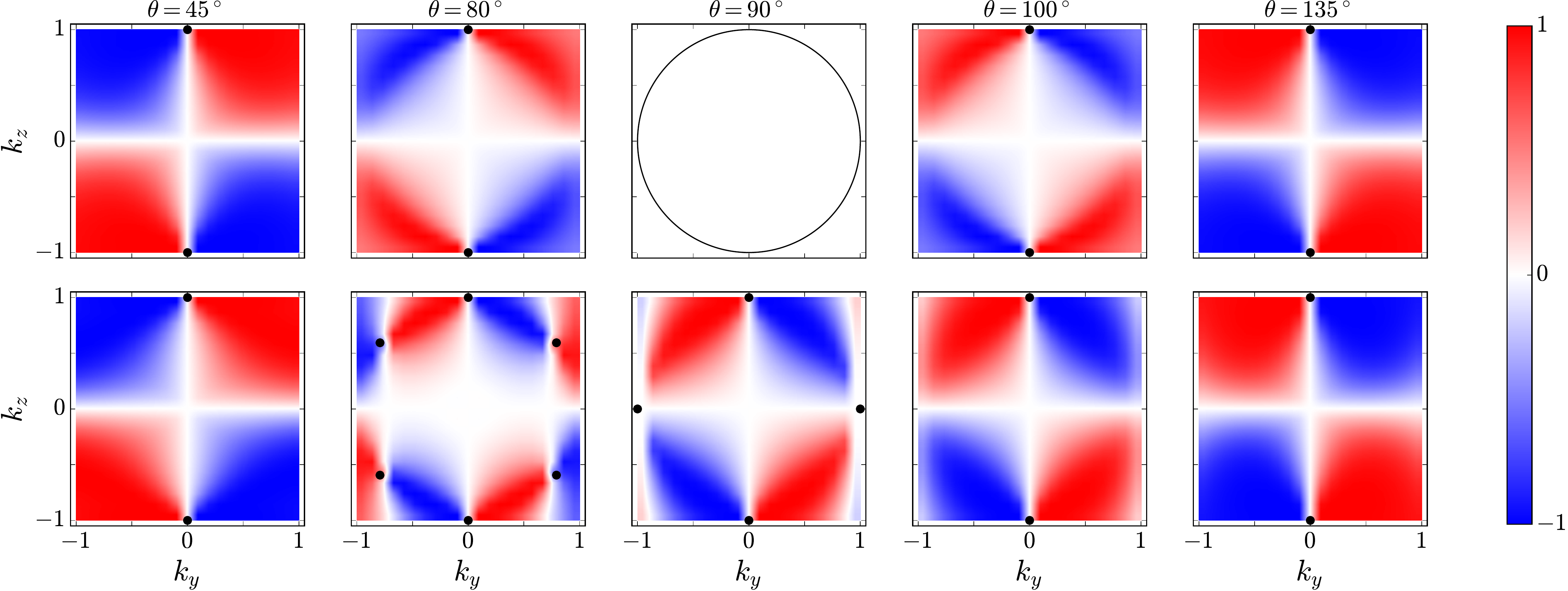}	\caption{Contour plot of circular dichorism, $\eta({\bf k})$ in $yz$ plane for type-II DSMs with (upper row) and without (lower row) mirror symmetry when the light is incident along the $y$-direction. The applied magnetic field is rotated in the $xz$ plane. Evidently, the overall pattern of $\eta$ differs from that of $\eta$  for light field along $x$-direction. In addition, at $\theta=\pi/2$, $\eta^y$ is zero everywhere except along the nodal line when mirror symmetry is present. In the absence of mirror symmetry, $\eta^y$ is zero at $\bf k=0$ at the mirror-symmetric angle. These are in contrast to the cases in Fig.~\ref{fig:eta-ii-x} where $\eta$ diverges at $\theta=\pi/2$ irrespective to the presence/absence of the mirror symmetry. Note that the nodal line is shown for guiding to eye. Parameters remain same as Fig. \ref{fig:eta-ii-x}}.	
	\label{fig:eta-ii-y}
\end{figure*}

The lower row of the Fig.~\ref{fig:eta-ii-x} illustrates the behavior of $\eta^x$ when mirror symmetry is broken. For smaller $\theta$, the pattern of $\eta^x$ is obtained to be similar to that of upper row of the Fig.~\ref{fig:eta-ii-x}. This can be understood easily from the Berry curvature as shown in Eq.~\ref{womirror_Berry}. The symmetry breaking term involving $\lambda$ contains $\sin(\theta)$, hence it does not modify $\Omega_x$ significantly for $\theta\ll \pi/2$. As $\theta$ increases, it starts contributing to $\Omega_x$, consequently we find different pattern in $\eta^x$. We note that $\eta^x$ diverges in the limit $(k_x, k_z) \rightarrow 0$ as before, but it is finite for non-zero $(k_x, k_z)$ at the mirror-symmetric angle. This is in contrast to the case which preserves mirror symmetry. Indeed, this fact can be used to identify whether a type-II DSM can host the mirror anomaly.

\subsection{Light perpendicular to the plane of magnetic field}
	
For incident light along $y$-direction and for \textit{mirror-symmetric} DSMs, we find $\eta^y$ in $k_x=0$ plane as 
\begin{equation}\label{eta_y_II_kx0}
\eta^{y}(k_y,k_z) = \frac{2 \epsilon_{\bm k} k_y k_z v \cos (\theta )}{\sqrt{k_y^2 \sin ^2(\theta )+k_z^2} \left(\epsilon_{\bf k}^2+\frac{k_z^2 v^2 \left(\epsilon_{\bf k}^2-p_-^2\right)}{(b-p_-)^2}\right)}.
\end{equation}
The circular dichroism $\eta^y$ as a function of $\theta$ for both the cases with and without mirror symmetry is depicted in Fig.~\ref{fig:eta-ii-y}. For any finite $\theta\ll\pi/2$, $\eta^y$ with mirror symmetry is found to be nearly zero in the $k_y$-$k_z$ plane, satisfying $k_y^2+k_z^2\ll b^2$  and on the other hand it is maximum above the line $k_y^2+k_z^2\ge b^2$ as evident from upper row of the Fig.~\ref{fig:eta-ii-y}. For the same range of $\theta$, the pattern differs from $\eta^x$.

\begin{figure}[h]
	\centering
	\includegraphics[width=0.48\textwidth]{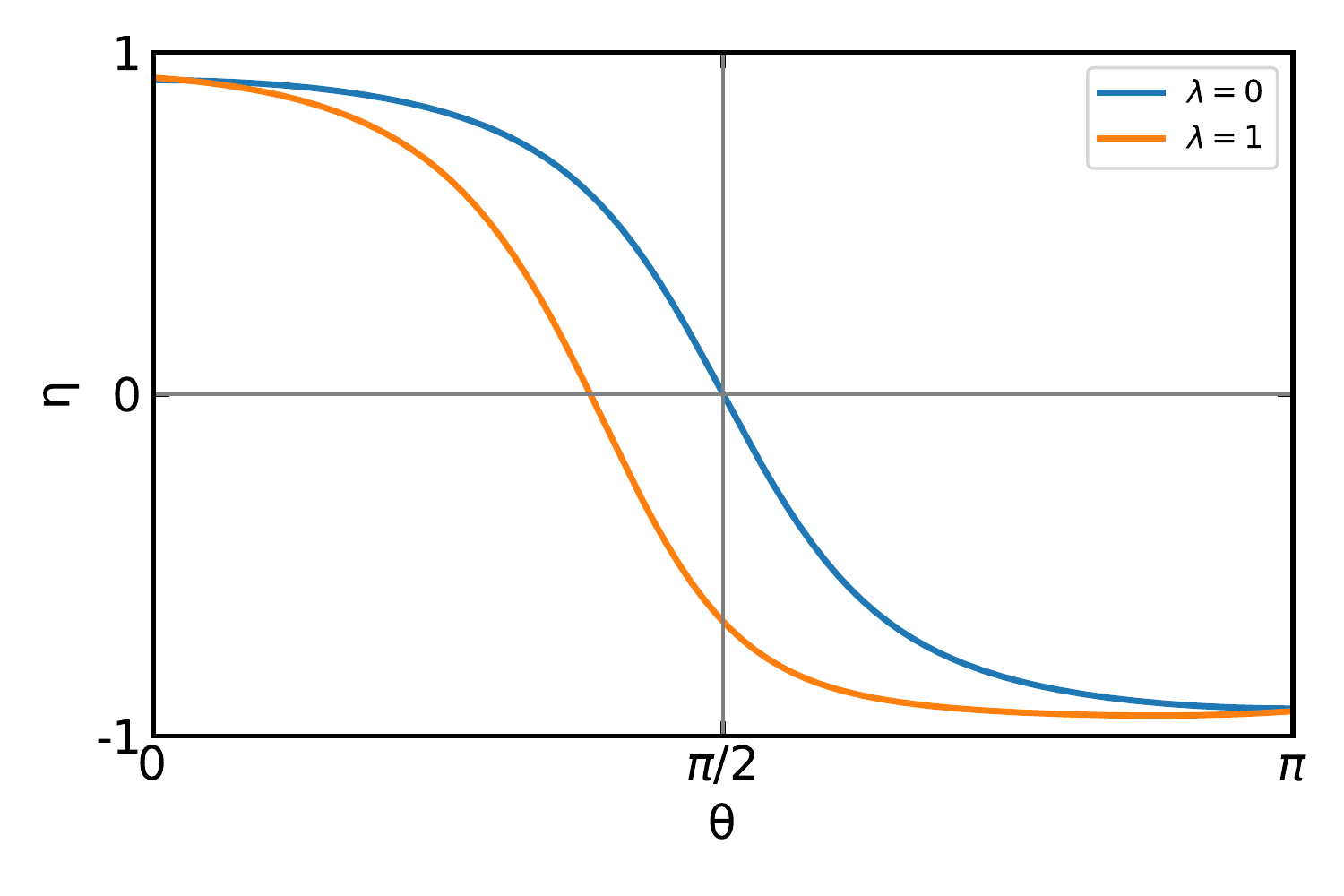}		\caption{Variation of circular dichroism ($\eta^y$) at finite momentum with respect to angle ($\theta$) of the external magnetic field rotated in $xz$ plane for type-II DSM with ($\lambda=0$) and without ($\lambda=1$) mirror symmetry. Here, the light is incident along $y$-direction. It is clear that for mirror-symmetric case, $\eta^y$ becomes zero at $\theta=\pi/2$ whereas picks up finite value when the mirror symmetry is broken. Parameters used: $ k_y=0.35, k_z=0.6 $ in units of $\AA^{-1}$.}	
	\label{fig:dsm_ii_eta-xz-y}
\end{figure}

This is also apparent from their corresponding equations (\ref{eta_y_II_kx0}) and (\ref{eta-x_ky0}). Notably, the crucial difference between $\eta^y$ and $\eta^x$ is found at the mirror-symmetric angle $\theta=\pi/2$. At $\theta=\pi/2$, $\eta^y$ is zero everywhere in the $(k_y,k_z)$ plane except at the nodal line, reflecting the vanishing Berry curvature at $\theta=\pi/2$. This in contrast to the $\eta^x$ which diverges due to the denominator although Berry curvature individually vanishes at this specific angle. 


\begin{figure*}
	\centering
	\includegraphics[width=0.99\textwidth]{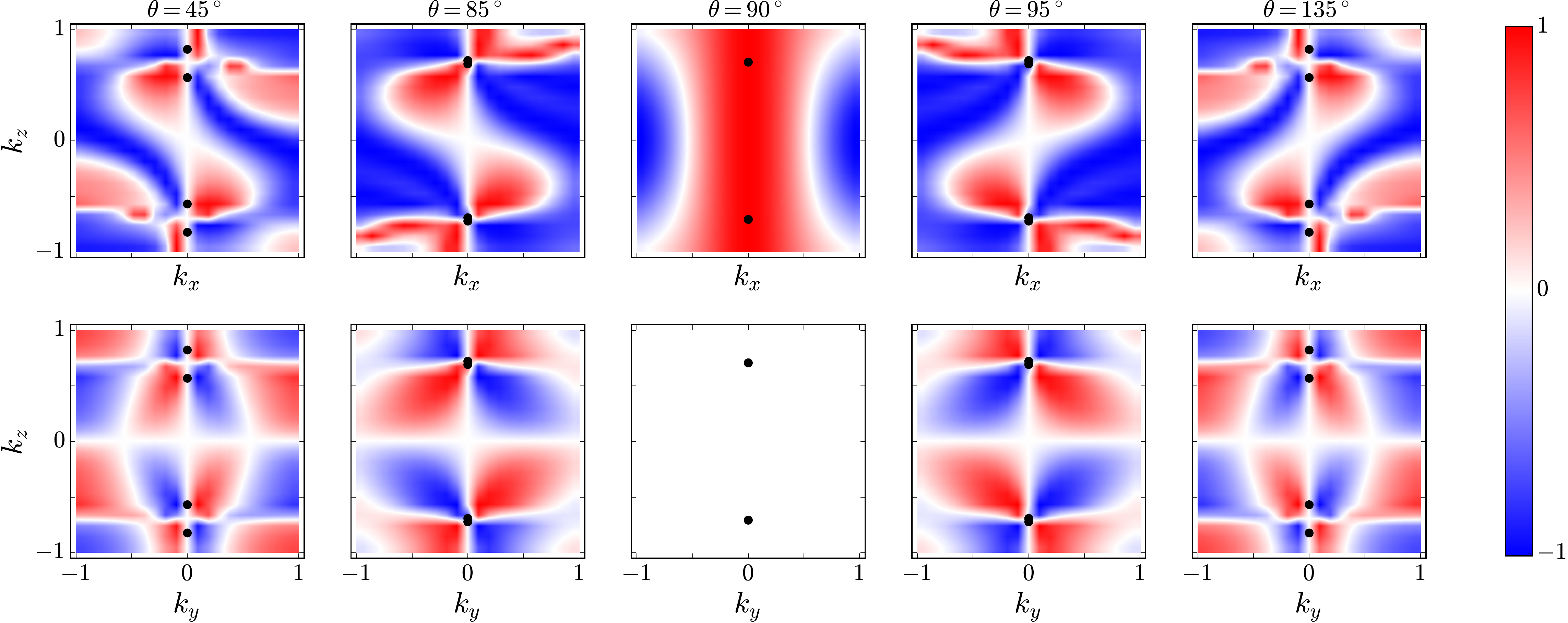}
	\caption{Contour plot of circular dichorism, $\eta({\bf k})$ in $xz$ plane and $yz$ plane for type-I DSMs with mirror symmetry considering the incident light along the $x$ direction ($\eta^x$ upper row) and $y$ direction ($\eta^y$ lower row) respectively. The applied magnetic field is rotated in the $xz$ plane. $\eta^x$ and $\eta^y$ differ from each other as clearly shown in figure. This distinction becomes sharp at mirror-symmetric angle $\theta=\pi/2$ where $\eta^y$ is found to be zero everywhere in the ($k_y$-$k_z$) plane in contrast to the $\eta^x$ which is finite except along two semi-circular lines in the $k_x-k_z$ plane. The Weyl nodes are marked with \textit{black dots}. The parameters are chosen as $ M_0 = -2A_1, M_{||} = -A_1/5, M_z = -4A_1 $ \cite{kobayashi_topological_2015}. Here we use $A_1/b=10$ in units of $\AA$.}
	\label{fig:dsm-i_xz_x.pdf}
\end{figure*}

For the case without mirror symmetry, $\eta^y$ shows similar pattern for $\theta\ll\pi/2$. However, as $\theta$ increases a complete contrasting feature is observed. Indeed this is due to the presence of six Weyl points for  $\theta>\theta_c$. For illustration we have indicated the location of Weyl points. At $\theta=\pi/2$, we find finite $\eta$ for some specific values of momentum in contrary to the case with mirror symmetry where $\eta^y$ is zero for any finite momentum except those on the nodal line. Thus, the effect of broken \textit{mirror symmetry} is well manifested in $\eta^y$, specifically at $\theta =\pi/2$ (cf. Fig. \ref{fig:eta-ii-y}). To justify this distinction, we also plot $\eta^y$ as a  function of $\theta$ in Fig.~(\ref{fig:dsm_ii_eta-xz-y}) for $k_y, k_z\ne0$. Note that this result also differs from $\eta^x$ without mirror symmetry, where it diverges at the zone center.

To close this section, we comment that in all the cases, the patterns of $\eta$ turn out to be some reflection symmetric about all the 3-axes $k_x$, $ k_y $ and $ k_z $ with polarities inverted. It is similar to a reflection in a plane followed by a rotation about an axis perpendicular to that plane, also known as \textit{rotoreflection}. In other words, $\eta$ turns out to be an odd function of each component of $\bm k$. It is important to note that although we have shown the behavior of $\eta(\mathbf{k})$ for $b/v=1$, the qualitative features of $\eta(\mathbf{k})$ remain unchanged for the range $b/v = 0.001-1$ with the typical value $v=10^5 m/s$ which corresponds to the strength of the magnetic field 1 to few hundreds Tesla and therefore, achievable in the experiments.

\section{Circular Dichroism in Type-I Dirac Semimetal}\label{type-i_dirac_semimetal}
We now investigate the CD in type-I DSM to see if there is any distinction between these two types of DSMs. Since type-I DSMs do not process any mirror anomaly\cite{nandy_mirror_2019} in the anomalous Hall conductance as mentioned before, we only concentrate on the case with mirror symmetry. The low energy effective Hamiltonian for this type of DSMs near $\Gamma$ point is given by \cite{hashimoto_superconductivity_2016, cano_chiral_2017, pikulin_chiral_2016},
\begin{equation}\label{h-i}
	\begin{split}
	\mathcal{H}^{\textrm{I}}(\bm k) = M(\bm k) \tau_z \sigma_0 &+ A(\bm k) \tau_x \sigma_z + B(\bm k) \tau_y \sigma_0 \\
	&+ C(\bm k) \tau_x \sigma_x + D(\bm k) \tau_x \sigma_y,
	\end{split}
\end{equation}
where, $ M(\bm k) = M_0 - M_{||} (k_x^2+k_y^2) - M_z k_z^2 $ is the momentum dependent mass parameter, $ A(\bm k) = A_1 k_x $, $ B(\bm k) = -A_1 k_y $,  $ C(\bm k) = (\beta+\gamma)k_z(k_y^2-k_x^2) $ and  $ D(\bm k) = -2(\beta-\gamma)k_zk_yk_x $. The band parameters $ M_0, M_{||}, M_z, A_1, \beta$ and $ \gamma $ are material dependents and can be obtained from first-principles calculations. The degenerate Dirac points occur at $ (0,0,\pm \sqrt{\frac{M_0}{M_z}})$. Although the Hamiltonian in Eq.~\ref{h-i} has several mirror-symmetric planes such as (001), (100) and (110) \cite{hashimoto_superconductivity_2016}, we only focus on the $k_y=0$ plane for simplicity. As before, we consider a magnetic field in the $k_y=0$ plane as \cite{hashimoto_superconductivity_2016}
\begin{equation}\label{h-i_zmn}
	\mathcal{H}^{\textrm{I}}_{zmn}(\theta) = b\cos(\theta)\tau_0 \sigma_z+b\sin(\theta)\frac{(\tau_0+\tau_z)}{2}\sigma_x.
\end{equation}
Eq.~(\ref{h-i_zmn}) breaks time-reversal symmetry, which results in two pairs of  Weyl points at $ (0,0,\pm \sqrt{\frac{M_0\pm B_z}{M_z}})$, where $B_z$ is the $z$-component of the applied field.

For finite $\theta$, the Hamiltonian $\mathcal{H}^{\textrm{I}}(\bm k)+	\mathcal{H}^{\textrm{I}}_{zmn}(\theta)$ cannot be expressed in block diagonalized form by the simple similarity transformations used for type-
II DSMs. Thus we resort to the Eq. (\ref{cd-tm}) to calculate the CD. In Fig.~\ref{fig:dsm-i_xz_x.pdf}, we compare $\eta$ along two different directions, $x$ and $y$, of the incident light. As before, the pattern in $\eta^x$ and $\eta^y$ differs from each other. However, this distinction becomes sharp at $\theta=\pi/2$ where $\eta^y$ is found to be zero everywhere except at the Weyl points on the $k_y$-$k_z$ plane. This is in contrast to the $\eta^x$ which is finite except along two semi-circular lines on $k_x$-$k_z$ plane. 

To compare the pattern with the type-II DSMs, we only consider the case with mirror symmetry. Clearly, at $\theta=\pi/2$, $\eta^x$ for type-I is finite at the zone center, whereas it diverges for the type-II. This is indeed a sharp feature which can be experimentally verified. However, $\eta^y$ is almost similar for both the type-I and type-II. Finally,we note that the \textit{reflection}  symmetry is no longer present in type-I DSMs. Specifically, $\eta$ is neither an odd function of $k_x $ nor that of $ k_z $. However, it still has the \textit{rotoreflection} symmetry. Moreover, we find the patterns \textit{mirror-reflected} about $ z $-axis at $\theta=\pi/2$. Thus these differences can also be used to differentiate type-I DSMs from type-II DSMs.

\section{Conclusion}\label{conclusion}
In conclusion, the DSMs can be classified into type-I and type-II based on their implicit symmetries such as TRS, IS and crystalline uniaxial rotational symmetry. Although type-I and type-II DSMs differ in underlying symmetries, there are no as such observable properties that are significantly different in these two class of DSMs. Moreover, it has been argued that unlike type-I DSMs, the type-II DSMs in the presence of mirror symmetry can posses a novel anomaly, namely the mirror anomaly in addition to the celebrated chiral anomaly. Although it has been proposed that the anomalous Hall conductivity is one of the observable consequences to detect the presence of mirror anomaly, there is no concrete experimental evidence yet. Specifically, an easy reliable probe to distinguish between type-I and type-II DSMs based on their symmetries as well as the presence/absence of mirror anomaly in DSMs is still lacking.

To answer this, we study momentum-resolved circular dichroism $\eta(\bf{k})$ in these two types of DSMs in the presence of a rotating magnetic field applied in any one of the mirror-symmetric planes. Interestingly, we find that $\eta(\bf{k})$ shows significantly different patterns in both types of DSMs for the light field parallel or perpendicular to the plane of the applied magnetic field. For a specific direction of the incident light and at the mirror-symmetric angle, the CD pattern in type-II DSMs with mirror symmetry was found to show a distinct feature as compared to the case without mirror symmetry. Therefore, the presence or absence of mirror symmetry in type-II DSMs is well manifested in the CD, specially at the mirror-symmetric angle which may help finding type-II DSMs that host step-like Hall conductance as a function of magnetic field, in particular, the mirror anomaly. In addition, we find that for the light field incident along the $x$-direction, the CD at the mirror-symmetric angle diverges at the zone center in type-II DSMs, whereas it vanishes in type-I. Moreover, in type-II DSMs, the CD is obtained to be an odd function of individual components of momentum, whereas it differs in type-I.

Finally, we make a few comments regarding experiments. The typical photon energy for circular dichroism is of the order of eV. However, this may vary depending upon the band gap which sets the required photon energy for the optical transitions. In the current study, the gap is roughly of the order of 0.05 eV near k = 0 (e.g., in type-I) at finite magnetic field ($\sim$few Tesla). Thus one can use photon energy $\sim$50 meV or greater for optical transitions in experiment. However, the theoretical calculation of k-resolved CD defined in Eq.~(\ref{eta}) does not explicitly depend on the photon energy with the form of vector potential used herein. Nevertheless, the signatures presented in the current study are experimentally verifiable and can be used to distinguish different class of DSMs. Since type-II DSMs with mirror symmetry can have a transistor-like action, our study may contribute to potential technological applications by detecting mirror anomaly along with the application in optoelectronic devices.

\section*{Acknowledgments}
A.R.K. acknowledges MHRD, India for a research fellowship. G.P.D. acknowledges the Center for Computational Materials Science, Institute for Materials Research, Tohoku University for use of the MASAMUNE-IMR Supercomputing Facility for part of the calculations.

	\bibliography{citations}
\end{document}